\documentclass[journal=apchd5,manuscript=draft]{achemso}
\usepackage[T1]{fontenc}
\usepackage[english]{babel}
\usepackage{amsmath,amsthm,amssymb,latexsym,amsfonts,gensymb} 
\usepackage[printonlyused,nolist]{acronym} 
\usepackage{graphicx,float,epstopdf,hyperref} 
\usepackage{hyperref}

\usepackage{xcolor}
\usepackage[normalem]{ulem}

\graphicspath{{../figures}}

\title{Light-matter interaction between templated molecular layers and surface lattice resonances}

\author{Roland Sch\"afer}
\affiliation{Institute for Light and Matter, Universit\"at zu K\"oln, Greinstr. 4--6, 50939 K\"oln, Germany}

\author{Manuel Neubauer}
\affiliation{Institute for Light and Matter, Universit\"at zu K\"oln, Greinstr. 4--6, 50939 K\"oln, Germany}

\author{Klaus~Meerholz}
\affiliation{Institute for Light and Matter, Universit\"at zu K\"oln, Greinstr. 4--6, 50939 K\"oln, Germany}

\author{Klas~Lindfors}
\affiliation{Institute for Light and Matter, Universit\"at zu K\"oln, Greinstr. 4--6, 50939 K\"oln, Germany}
\email{klas.lindfors@uni-koeln.de}

\begin{document}

\maketitle

\begin{acronym}
		\acro{AOI}[AOI]{angle of incidence}
		\acrodefplural{AOI}{angles of incidence}
		\acro{PVD}[PVD]{physical vapor deposition}
		\acro{PL}[PL]{photoluminescence}
        \acro{TDM}[TDM]{transition dipole moment}
		\acro{GNR}[7-AGNRs]{seven-atom wide armchair-edge graphene nanoribbons}
		\acro{SLR}[SLR]{surface lattice resonances}
		\acro{LSPR}[LSPR]{localized surface plasmon resonances}
		\acro{DO}[DO]{diffraction order}
		\acro{NA}[NA]{numerical aperture}
	\end{acronym}
	
\begin{abstract}
We couple a templated layer of merocyanine molecules with surface lattice resonances in a plasmonic grating. The templating of the molecular layer is achieved using a layer of aligned graphene nanoribbons, resulting in anisotropic optical properties. The anisotropy manifests itself in polarization-dependent coupling between excitons in the organic layer and lattice plasmons in the grating. We study the influence of the templating on the coupling and find that surprisingly the more orientational ordered templated layer displays a lower coupling strength than an identical amorphous merocyanine layer. Our work demonstrates the use of molecular templating in controlling the interaction of excitons with excitations in optical nanostructures.
\end{abstract}

\section*{Introduction}

Confining light to small mode volumes over an extended amount of time increases the coupling between light and matter. If the interaction strength is large enough, the strong coupling regime is reached. In this regime, the light-matter interaction exceeds the
total losses of the system and the re-absorption of photons by the material becomes the dominant process. Here, hybrid light-matter states are formed, called polaritons.\cite{GarciaVidal2021} Polaritons have properties of both the light and the matter component, which has lead to interesting observations in recent studies, such as improved energy transport in transistors\cite{Dutta2024,Bhatt2021,Orgiu2015}, lasing at room temperature\cite{Daskalakis2014} and modified chemical reaction rates\cite{Hutchison2012,Thomas2016}. Systems that confine light and materials with large oscillator strength are required to reach the strong coupling regime. Strong coupling has been demonstrated in a multitude of geometries, such as planar microcavities, micropillars\cite{Gerard1998}, microdisks\cite{Gayral1999} and photonic crystals\cite{Srinivasan2003}. In recent years, plasmonic lattices have gained interest in studying light-matter interaction due to the formation of \ac{SLR}. These are sharp modes that form by a collective coupling between nano-particles ordered in an array with a lattice period on the order of magnitude of the wavelength of the light. Due to this collective coupling, the photons are confined which results in an increased light-matter coupling. This has been used to study strong coupling\cite{Vaekevaeinen2013}, long-range energy propagation\cite{Yadav2020}, and for biosensing applications\cite{Gutha2017}.

Strong coupling between plasmonic lattices and organic molecules has mostly been studied with disordered materials.\cite{Vaekevaeinen2013,Zhou2013,Yang2015,Wang2017} Recently, it was shown for planar microcavities that molecular alignment can lead to a larger coupling strength, which is a important parameter for, e.g.\@, polariton lasing.\cite{Schaefer2025} For plasmonic lattices, it has been shown that an anisotropic single crystal of tetracene can be used to control the light-matter coupling strength from the weak- to the strong-coupling regime.\cite{Berghuis2020} Ordered excitonic components are thus highly interesting for studies of light-matter coupling in plasmonic lattices.

In this work, we investigate the influence of order on coupling between plasmonic lattices and ordered molecular layers. We study both the influence of dynamic and static disorder. Dynamic disorder results in broadening of the absorption line by overlap of the exciton and molecular vibrations in molecular crystals, while static disorder describes, e.g.\@, the orientation of molecules (orientational disorder). Here, we use the merocyanine dye 2-[5-(5-dibutylamino-thiophen-2-yl-methylene)-4-\textit{tert}-butl\-5\textit{H}\-thiazol\-2-ylidene]\-malononitrile (HB238\cite{Buerckstuemmer2011}) as a model system. Merocyanines are a class of molecules that tend to form molecular aggregates.\cite{Wuerthner2011} The aggregation can result in a red-shifted (J-aggregate), blue-shifted (H-aggregate), or the appearance of both a red- and a blue-shifted absorption relative to the monomer. J- and H-aggregates are formed by `head-to-tail' and `side-by-side' arrangement of the molecules, respectively.\cite{Kasha1963,Kasha1965,Wuerthner2011,Hestand2018} In the case of HB238, the molecules aggregate in oblique-angle configuration, resulting in both a J- and H-like spectral transition.\cite{boehnerphd,Schaefer2023,Boehner2025} In a recent publication we have shown that both the J- and the H-transition can be strongly coupled in planar microcavities (see Refs.~\cite{Schaefer2023,Schaefer2025}). In amorphous (spin-cast) thin films of HB238, the \ac{TDM} of the J-transition is parallel to the sample plane, but has no defined direction. Here, we use \ac{GNR} as a template to orient the \ac{TDM} of the J-transition to study and control the light-matter interaction with plasmonic lattices.  

\section*{Results and discussion}

\begin{figure}[H]
    \centering
    \includegraphics[width=0.95\textwidth]{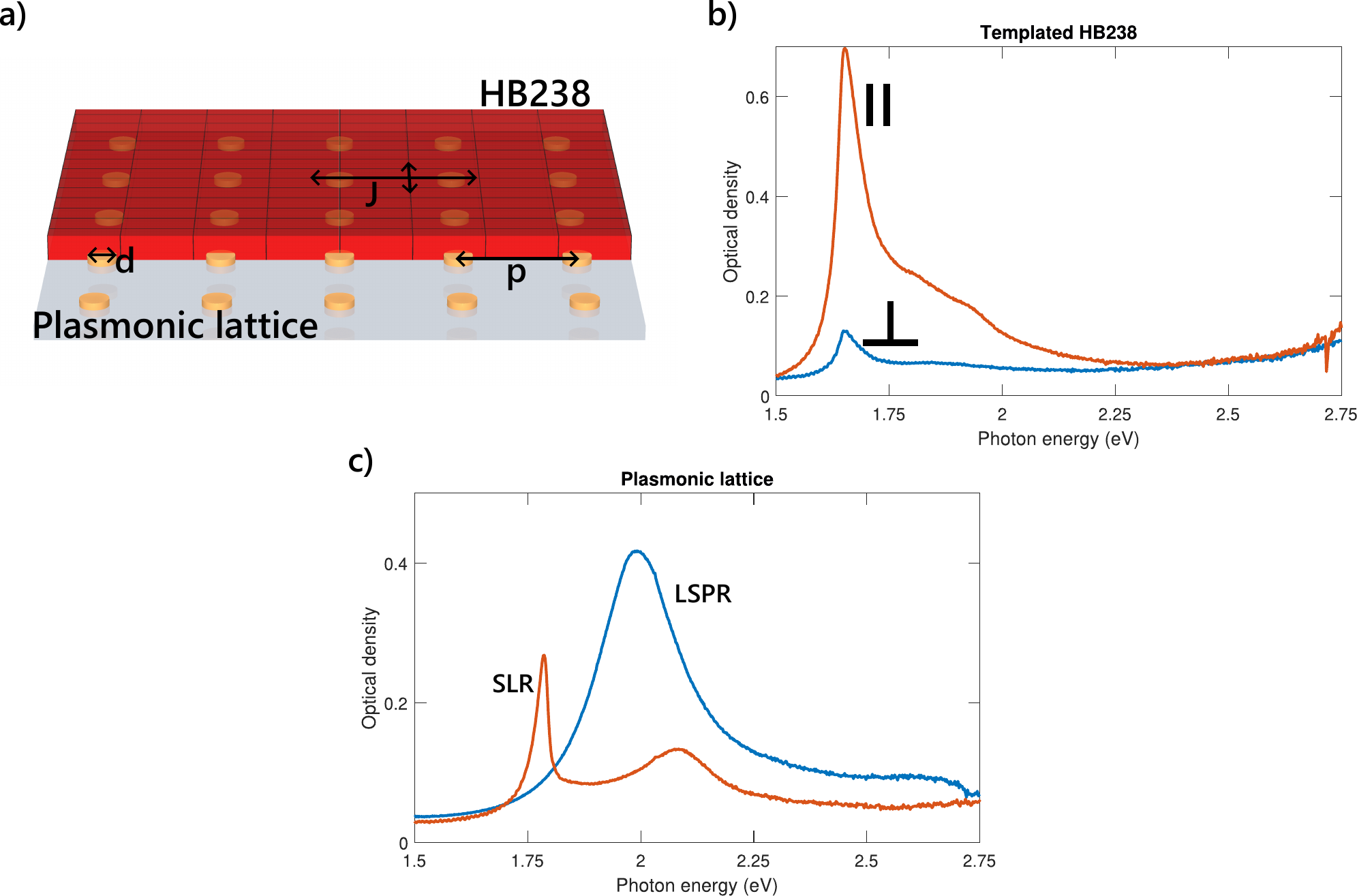}
    \caption{a) Sketch of the investigated system: Plasmonic lattices with particle spacing p and diameter d. An aligned layer of HB238 is deposited on the lattice. The molecular alignment is achieved by templating by graphene nanoribbons. b) Transmission spectra of the templated HB238 film polarized parallel ($\parallel$) and perpendicular ($\perp$) to the graphene nanoribbons. c) Transmission spectra of \acl{LSPR} mode from a subdiffractive lattice (period: 300~nm) and of a \acl{SLR} mode.
    }\label{fig:SLR:resarch_idea}
\end{figure}

We investigate the light-matter interaction between a plasmonic gold lattice that supports \ac{SLR} modes with aligned (templated) molecular layers of HB238. The system is sketched in Fig.~\ref{fig:SLR:resarch_idea}a.  In short, the molecule HB238 belongs to a group of merocyanines that have been studied due to their ability to form aggregates in thin films displaying J- or H-transitions.\cite{Buerckstuemmer2011,Liess2017,Liess2018} HB238 molecules aggregate in oblique-angle fashion, resulting in both a J- and H-like transition, with the \ac{TDM} of the J- and H-transition aligned parallel and perpendicular to the substrate plane, respectively. Due to the fascinating physical properties properties, HB238 has been intensively studied.\cite{Tomar2023,Gildemeister2021,boehnerphd,Boehner2025, weitkampphd,Schaefer2023,Schaefer2025,Oecal2025,Kny2023} Weitkamp\cite{weitkampphd} demonstrated that the molecular crystals of HB238 can be aligned by physical vapor deposition on \ac{GNR}\@. The \ac{GNR} serve as a template for the molecular aggregates, inducing uniaxial growth parallel to the long axis of the \ac{GNR}.\cite{weitkampphd} This results in a highly polarization dependent response: if the electric field of light is parallel to the \ac{GNR}' long axis, the absorption of the J-transition at 1.65 eV is observed as strong as shown in Fig.~\ref{fig:SLR:resarch_idea}b. The signal strongly decreases, if the sample or polarization is rotated by 90$^\circ$ (see Fig.~\ref{fig:SLR:resarch_idea}b). Note that the \ac{TDM} of the H-transition is parallel to the substrate normal and therefore not accessible at normal incidence. For the templated layer, the H-transition can still be observed at non-normal \ac{AOI} and p-polarized light (see Refs.~\cite{Schaefer2023,Schaefer2025}). 

In Ref.~\cite{Schaefer2025} we studied the influence of the alignment resulting from templating on strong light-matter interaction in a planar microcavity. The molecular alignment results in the observation of polarization splitting of the polaritons, with increased Rabi splitting compared to a non-templated film, if the electric field component of light is parallel to the \ac{GNR} long axis. In the perpendicular case, the coupling strength is minimized. If the cavity is rotated by any other angle, all four polaritons are observed. This polarization splitting allowed us to spatially probe the quality of alignment  and the influence of order on light-matter interaction.\cite{Schaefer2025} Here, we investigate the coupling between \ac{SLR} and templated HB238. The \ac{SLR} are the collective coupling between \ac{LSPR} of plasmonic nano-particles in a periodic lattice with the \ac{DO} of the array, resulting in sharp resonances with large $Q$-factors as illustrated in  Fig.~\ref{fig:SLR:resarch_idea}c. The \ac{SLR} allow for a multitude of geometries such as square, rectangular, hexagonal\cite{Guo2017} or superlattices\cite{Wang2015} with tailored light fields, which makes \ac{SLR} an interesting platform to study light-matter interaction with ordered molecular layers. In literature, disordered molecules\cite{Vaekevaeinen2013,Zhou2013,Yang2015,Wang2017} are typically used for light-matter studies involving plasmonic lattices. Recently, Berghuis \textit{et al.}\cite{Berghuis2020} investigated the interaction between a silver nano-particle array and a tetracene single crystal, which has anisotropic optical properties. The study includes the rotation of the crystal on the lattice, which allows for controlling the coupling strength from weak to strong, depending on the rotation angle.\cite{Berghuis2020} 

\begin{figure}[H]
    \centering
    \includegraphics[width=0.95\textwidth]{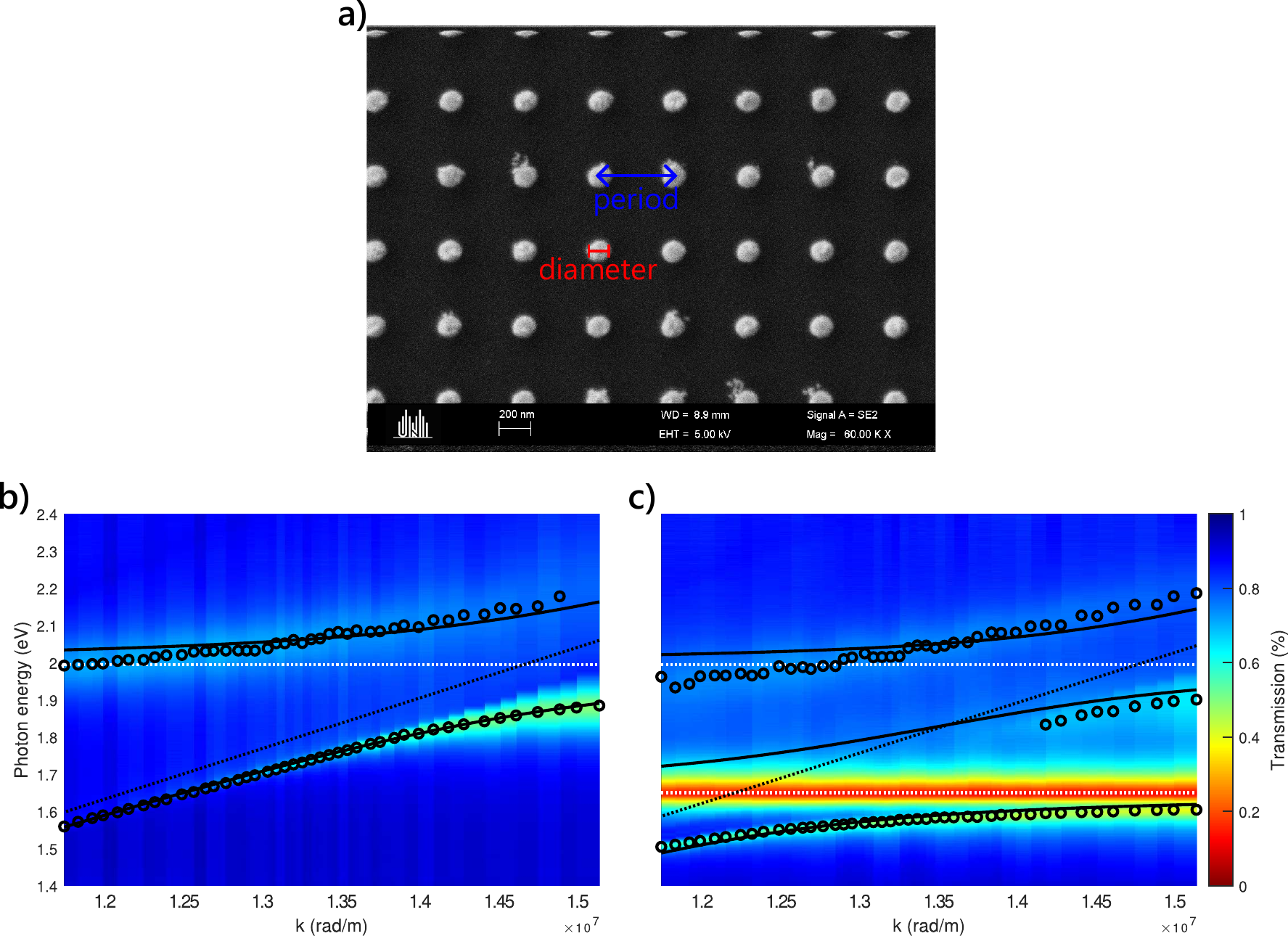}
    \caption{a) Scanning electron micrograph of a gold nano-particle lattice with a diameter of approximately 92~nm and a period $p$ of 475~nm. b) and c) Transmission spectra as a function of the inverse lattice spacing $k=2\pi/p$ of lattices covered with polyvinyl alcohol and lattices  coupled to HB238, respectively. Black circles indicate the measured spectral peak positions. The horizontal white dotted lines at approximately 2~eV and 1.65~eV is the \acl{LSPR} of the gold particles (panel b and c) and the J-transition of HB283 (panel c), respectively. The black dotted line is the calculated position of the \acl{DO} (see Equation~\ref{eq:SLR:do}) and the black solid lines are the fit results from the coupled oscillator models, Equation~\ref{SLR:fit:bare} and \ref{SLR:fit}. 
    }\label{fig:SLR:spin_cast}
\end{figure}

The plasmonic array consists of gold nano-disks with a diameter of 92~nm and a height of 50~nm. Arrays with different periods were studied in order to shift the \ac{SLR} spectrally. We first characterize the lattice without the excitonic layer of HB238. A homogeneous refractive index environment is achieved by spin-casting a  few hundred nanometers thick layer of polyvinyl alcohol (see Methods for full details), which is required for sharp \ac{SLR} modes at normal incidence (see Ref.~\cite{Kravets2018}). To detune the photonic mode, we fabricated lattices with periods in the range of 415-535~nm. A representative scanning electron micrograph of an array is shown in Fig.~\ref{fig:SLR:spin_cast}a. In Fig.~\ref{fig:SLR:spin_cast} we display the transmission spectra as a function of the inverse lattice spacing $k=2\pi/p$, where $p$ is the (real space) lattice period, following the approach of Väkeväinen \textit{et al.}\cite{Vaekevaeinen2013}\@. Clear anti-crossing is observed between the \ac{LSPR} and \ac{DO}\@. From now on the sharper and lower energy signal of the two formed modes will be referred as \ac{SLR}. The coupling can be described by a two-oscillator Hamiltonian\cite{Vaekevaeinen2013,Rodriguez2013}
\begin{equation}\label{SLR:fit:bare}
    H =     \begin{pmatrix}
        E_{\mathrm{LSPR}} & g_{1}\\
    			g_{1} & E_{\mathrm{DO}} 
    \end{pmatrix},
\end{equation}
with the \ac{LSPR} energy  $E_\mathrm{LSPR}$ and the \ac{DO} energy at normal incidence $E_\mathrm{DO}$, respectively. The interaction strength is given by the coupling parameter $g_1$. $E_\mathrm{DO}$ is dependent on the effective refractive index of the medium $n_{\mathrm{eff}}$ and the lattice period as
\begin{equation}\label{eq:SLR:do}
    E_{\mathrm{DO}} = \frac{\hslash\mathrm{c}}{n_{\mathrm{eff}}}\frac{2\pi}{p}.
\end{equation}
The fit by diagonalizing the Hamiltonian yields a coupling strength of approximately 130~meV between the \ac{DO} and the \ac{LSPR}. Note that the coupling parameter $g_1$ is half of the Rabi energy ($\hslash\Omega = 2g_1$).

Now we turn to the investigation of light-matter interaction between the \ac{SLR} and spin-cast HB238 (see Fig.~\ref{fig:SLR:spin_cast}). Note that a few hundred nanometers thick layer of polyvinyl alcohol was spin-cast on the HB238\@, because the approximate 20~nm HB238\cite{Schaefer2023,Boehner2025} layer is not sufficient to achieve a homogeneous refractive index environment to form sharp \ac{SLR} modes at normal incidence. In this case, four signals are observed: one period-independent peak at 1.65~eV, which corresponds to the J-transition of HB238\@, indicating that not all aggregates are coupled or only weakly coupled to the \ac{SLR} mode. The other three signals show dispersion with respect to the lattice spacing and display clear anti-crossing behavior, indicating strong light-matter interaction. Three new eigenstates have formed by the interaction between the \ac{LSPR}, \ac{DO} and HB238\@: a lower, middle and upper polariton.  We use a three coupling oscillator model\cite{Vaekevaeinen2013}, in order to quantify the coupling strength:
\begin{equation}\label{SLR:fit}
    H =     \begin{pmatrix}
        E_{\mathrm{LSPR}} & g_{1} & g_{2}\\
    			g_{1} & E_{\mathrm{DO}} & g_{3}\\
    			g_{2} & g_{3} & E_{\mathrm{HB238}}
    \end{pmatrix},
\end{equation}
where $g_{2}$ and $g_{3}$ quantify the coupling between the \ac{LSPR} and the J-transition (with energy $E_{\mathrm{HB238}}$) and the coupling of the \ac{DO} and J-transition, respectively. A fit of the eigenenergies obtained by diagonalizing the Hamiltonian Equation~\ref{SLR:fit} to the data shows good agreement (see Fig.~\ref{fig:SLR:spin_cast}c). The best fit yields approximately 105~meV for $g_{1}$ (\ac{SLR} mode) and 112~meV for $g_{3}$ (coupling of HB238 to the \ac{SLR} mode). The parameter $g_{2}$ is close to 0~meV, due to the large energy difference between the \ac{LSPR} ($\approx2$~eV) and HB238 ($1.65$~eV). Therefore, the dispersion of the upper polariton is shifted to similar energies as in the case without HB238 (compare Fig.~\ref{fig:SLR:spin_cast}b and c).

\begin{figure}[H]
    \centering
    \includegraphics[width=0.80\textwidth]{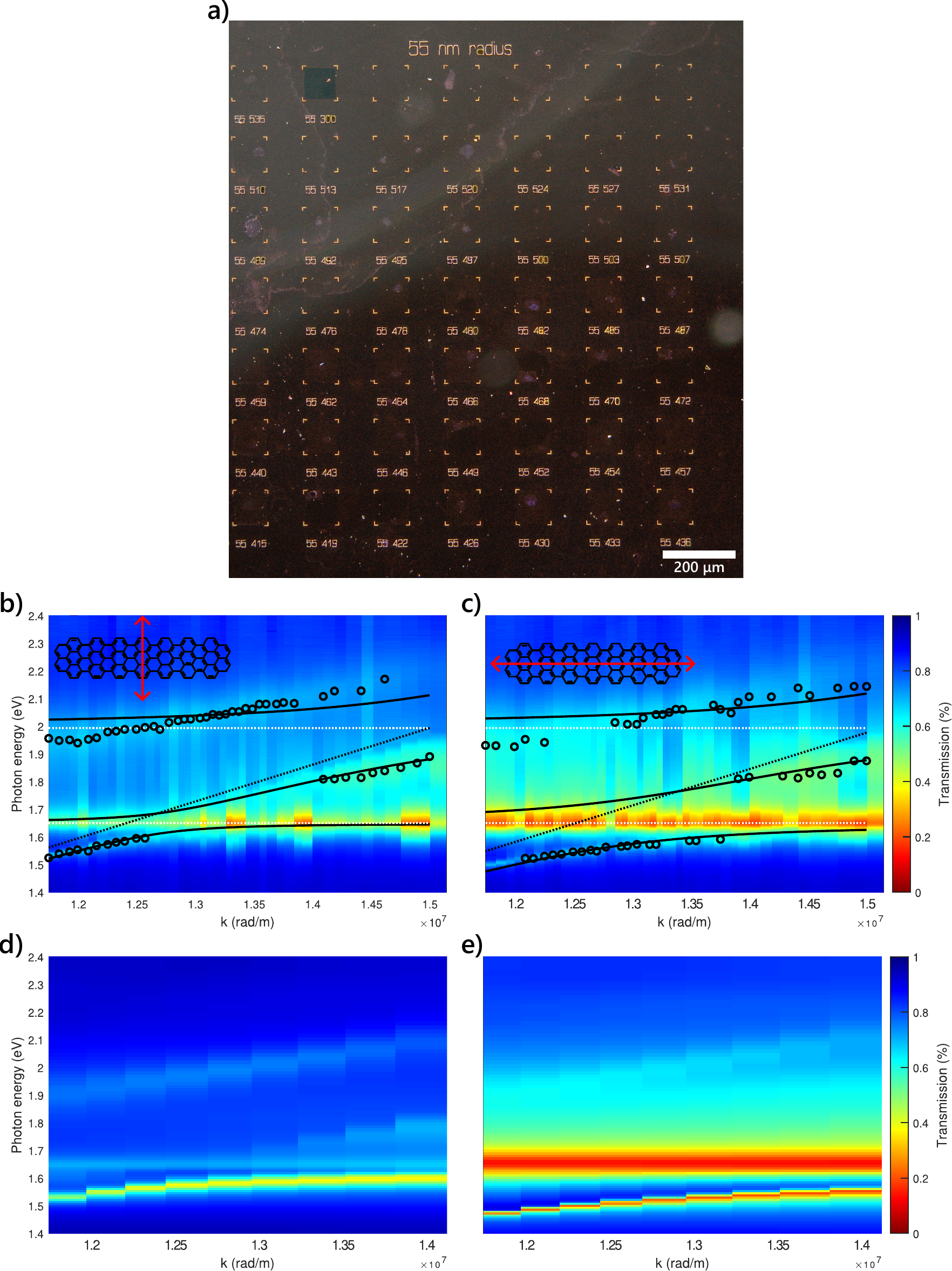}
    \caption{a) Dark-field microscope image of gold lattices with a templated HB238 layer. The nominal radius is 55~nm, which resulted in disks with a diameter of 92~nm. The second number under each lattice index is the lattice period. b), c) Measured transmission spectra of gold nano-particle lattices coupled with a templated HB238, for the electric field perpendicular (panel b) and parallel (panel c) to the graphene nanoribbons.  The horizontal white dotted lines at approximately 2~eV and 1.65~eV are the \acl{LSPR} of the gold particles and the J-transition of HB283, respectively. The black dotted line is is the calculated position of the \acl{DO} (see Equation~\ref{eq:SLR:do}) and the black solid lines are the fit results from three coupled oscillators model. d), e) Simulated transmission spectra of gold nano-particle lattices with templated HB238. $k$ is the inverse lattice period ($k=2\pi/\mathrm{period}$). 
        }\label{fig:SLR:templated}
\end{figure}

We have thus demonstrated that the strong coupling regime can be reached by the interaction between \ac{SLR} and spin-cast HB238. We now continue exploring the light-matter interaction with templated HB238 layers.  The oriented TDM of the transitions in templated layers offers much potential for engineering the light-matter interaction in SLRs. A dark-field microscope image of the sample is shown in Fig.~\ref{fig:SLR:templated}a, displaying a homogeneous layer of HB238, with the exception at the top left of Fig.~\ref{fig:SLR:templated}a, where cracks and defects can be seen. 

The \ac{GNR} were carefully aligned parallel to one of the lattice edges. In the spectroscopic experiments the sample was aligned such that the horizontal polarization is perpendicular to the \ac{GNR}.  Fig.~\ref{fig:SLR:templated}b shows the transmission spectra for horizontally polarized light. In this case, the electric field is perpendicular to the J-transition of HB238 (see Fig.~\ref{fig:SLR:resarch_idea}b). Even though the absorption of the J-transition is minimized in this configuration, the light-matter interaction is large enough to reach the strong coupling regime. The coupling strength between the \ac{SLR} and HB238 was determined as approximately 39~meV, by using the three coupled oscillator model (Equation~\ref{SLR:fit}). 

If the electric field is polarized parallel to the \ac{GNR}, the light-matter interaction increases to approximately 88~meV\@. Interestingly, the coupling strength is only doubled, while the extinction ratio (determined from the peak maxima) in the thin film is almost a factor of six between the two polarizations (see Fig.~\ref{fig:SLR:resarch_idea}b). Also in the parallel case, the coupling strength is lower than if HB238 is spin-cast (compare Fig.~\ref{fig:SLR:spin_cast}c and Fig.~\ref{fig:SLR:templated}c). For a qualitative comparison, finite element simulations were employed using the complex biaxial refractive index of HB238, which takes into account the effect of the templating.
(see Fig.~\ref{fig:SLR:templated}d and e). Both the perpendicular (Fig.~\ref{fig:SLR:templated}d) and parallel (Fig.~\ref{fig:SLR:templated}e) case agree well with the experimental data: The lowest polariton clearly displays anti-crossing behavior with the HB238, while displaying a similar dispersion strength as the experimental data, indicating similar coupling strength. The reduced coupling strength in going from spin-cast to templated HB238 film may be a result of aggregates with a low number of coherently coupled molecules which lead to a broadening of the J-transition (see Fig.~\ref{fig:SLR:resarch_idea}b and Ref.~\cite{Schaefer2025}). Therefore, not all oscillator strength lies in the J-transition, ultimately reducing the number of aggregates coupled to the \ac{SLR}. This result is in stark contrast to the increased light-matter interaction between planar microcavitites and templated HB238 in Ref.~\cite{Schaefer2025} due to the molecular alignment. It seems that in planar microcavities the static disorder plays a larger role for the light matter interaction, while dynamic disorder dominates the coupling strength in plasmonic lattices. Further studies may reveal a more detailed picture of the optical processes involved. 

Surprisingly, the middle polariton is not clearly visible in Fig.~\ref{fig:SLR:templated}e, while the other spectral features agree well with Fig.~\ref{fig:SLR:templated}c. This may also be a result of the broadened J-transition in the templated HB238 film: Because not all aggregates are strongly coupled to the \ac{SLR}, the J-transition is visible in transmission spectra of HB238 films on plasmonic lattices (in contrast to the results with microcavities in Refs.~\cite{Schaefer2023,Schaefer2025}). The spectra of the uncoupled or weakly coupled aggregates therefore overlap with the middle polariton, which results in reduced visibility of the middle polariton. 

\section*{Conclusion}

To conclude, we investigated the light-matter interaction between plasmonic lattices and HB238. HB238 aggregates in oblique-angle configuration, with the \ac{TDM} of the J-like transition parallel to the substrate plane and the \ac{TDM} of the H-like transition parallel to the substrate normal. Therefore, only the J-transition is optically accessible with measurements at normal incidence. Molecular alignment with \ac{GNR} as a template results in a highly polarization dependent response, with strong absorption at 1.65~eV, when the electric field component of light is parallel to the alignment direction. The intensity drops by a factor of almost six, when the electric field is polarized perpendicular to this. These optical properties make (templated) HB238 an interesting system to investigate strong light-matter interaction. In this study we used plasmonic lattices to confine light to a small mode volume for an extended amount of time to reach the strong coupling regime. 

Plasmonic lattices support \ac{SLR} modes by coupling the \ac{DO} of the lattice with the \ac{LSPR} of plasmonic nano-particles. We fabricated gold nano-disk lattices with a \ac{LSPR} mode at approximately 2~eV and periods in the range of 415-535~nm.  The resulting \ac{SLR} mode lies in an energy range to couple to the J-transition of HB238. By spin-casting HB238 (\ac{TDM} of the J-transition randomly oriented in the lattice plane), the strong coupling regime was reached with a coupling strength of approximately 112~meV. 

With molecular alignment by \ac{GNR}, we expect reduced and increased coupling strength when exciting molecules parallel and perpendicular to the alignment direction, respectively. Instead, the light-matter interaction was reduced in both cases, compared to the spin-cast case. If the electric field component and the aligned molecules are perpendicular to each other, the coupling strength is approximately 39~meV. In the parallel case, the coupling strength reaches a value of approximately 88~meV. The polarization dependence of the coupling strength confirms molecular alignment, but the overall reduction of the light-matter interaction might be caused by increased dynamic disorder. The increased disorder can be observed by the broadened J-transition in the thin-film transmission spectrum. Finite element simulations reproduce the experimental data well. Due to these unexpected results, the light-matter interaction between templated molecules and \ac{SLR} seem non-trivial and requires further investigations with, e.g.\@, other lattice geometries, such as anisotropic nano-particles, rectangular, or hexagonal lattices. The results may be used to further study the molecular order. Getting a better understanding of the molecular order may make molecular templating a feasible method to tune the optical properties for organic optoelectronics. The strong polarization dependent response of the coupling strength is already a promising property for polarization-sensitive applications such as polarizers. 

\newpage
\section*{Methods}

\subsection*{Fabrication of plasmonic lattices}

Plasmonic lattices were fabricated by electron beam lithography on borosilicate glass substrates (D263 T, \emph{Präzisions Glas \& Optik GmbH}). The structures were exposed with an electron beam with a 225~$\mu$C/cm$^2$ dose on a double layer of poly(methyl 2-methyl\-propenoate) (AR-P 642.04 and AR-P 672.02, \emph{Allresist}). To prohibit charging, a layer of an electron conducting polymer (ESpacer 300Z, \emph{Showa Denko}) was deposited on the resist. The bottom layer (AR-P 642.04, \emph{Allresist}) has a thickness of 150~nm, followed by a second layer of 50~nm (AR-P 672.02, \emph{Allresist}). The bottom layer's electron sensitivity is larger than that of the top layer, resulting in a undercut after development in a methyl isobutyl ketone/isopropanol solution (1:3 volume ratio). 3~nm of titanium and 50~nm of gold was subsequently evaporated via thermal physical vapor deposition. The metal on the polymer was removed with a lift-off procedure in dimethylsuccinat (AR 300-76, \emph{Allresist}).  Lattices  with periods from 415 to 535~nm and 300~nm were fabricated. The size for all lattices was 90~$\mu$m~$\times$~90~$\mu$m. For a homogeneous refractive index around the particles, an approximately 320~nm thick layer of polyvinyl alcohol was spin-cast from a water solution on either the plasmonic lattice or HB238.\\[6pt]

\subsection*{Spin-cast HB238 layer}

100~$\mu$l of HB238/Chloroform solution with a concentration of $1\times10^{-2}$~mol/l was spin cast on the plasmonic lattice, yielding a thin layer of approximately 20~nm.\cite{Schaefer2023,Boehner2025} The substrate was then annealed at 140~$^\circ$C to induce aggregation of the molecules.

\subsection*{Synthesis and transfer of \acl*{GNR}}

For the synthesis and transfer of \ac{GNR} the same protocol as in Ref.~\cite{Schaefer2025} was used: \ac{GNR} were synthesized on a clean Au(788) single crystal in ultra high vacuum by deposition of 16~\AA{} of 10,10’-dibromo-9,9’-bianthryl and subsequent heating to 200 and 400~$^\circ$C. The \ac{GNR} film was transferred from the gold crystal to the plasmonic lattice sample, by first covering the \ac{GNR} layer with poly(methyl 2-methyl\-propenoate) and delaminating the \ac{GNR} with the poly(methyl 2-methyl\-propenoate) by the bubble transfer method\cite{Sun2016}. The delaminated film was washed trice, and then placed on the plasmonic lattice substrate while carefully aligning the \ac{GNR} parallel to a lattice axis and dried for 24~h. The sample was subsequently placed in boiling acetone to remove the poly(methyl 2-methylpropenoate) layer.

\subsection*{HB238 deposition on 7-AGNRs}

A 25~nm thick layer of  the molecule HB238 was deposited with thermal physical vapor deposition with a rate of 0.03~\AA/s while heating the substrate to 74~$^\circ$C under high vacuum conditions. Atomic force microscope images (MFP-3D Infinity AFM, \emph{Oxford Instruments Asylum Research}) were recorded in alternating contact mode using AC200TS tips (\emph{Olympus}) to investigate the alignment. The atomic force micrographs were analyzed using the Gwyddion software\cite{Necas2012}.

\subsection*{Transmission spectra}

Transmission spectra were recorded by illuminating the sample with a halogen lamp from the back of the substrate. The light is collected with a 4x microscope objective (UPlanFL N, \emph{Olympus}) with a \ac{NA} of 0.13. Therefore, the objective only collects small angles. The transmitted light is analyzed with a broadband linear polarizer (UBB01A, \textit{Moxtek}) and recorded by focusing the light on the spectrometer entrance slit of an imaging spectrometer (IsoPlaneSCT 320 with deep cooled Pixis camera, \textit{Princeton Instruments}). The arrays were selected by using the imaging capabilities of the spectrometer and incrementally moved by the motorized stage to the next array. Multiple spectra from each lattice were averaged to increase the signal to noise ratio. Reference spectra were measured with the sample removed.

\subsection*{Data analysis}

The transmission spectra were analyzed and visualized with Matlab (\emph{MathWorks}). First multiple spectra from the same lattice were averaged to achieve a better signal to noise ratio, followed by dividing the measurements with the spectrum of the halogen lamp (measured without sample). The minima in the transmission spectra were identified with a peak finding function, and the minima were subsequently fitted with a two- or three oscillator model using the trust-region-reflective algorithm. A subdiffractive lattice (lattice spacing 300~nm) was used to determine the \ac{LSPR} spectral position.

\subsection*{Finite element simulations}

The finite element simulations were performed with the Wave Optics module in frequency domain (COMSOL Multiphysics, \emph{COMSOL Multiphysics GmbH}) to simulate transmission spectra. A gold cylinder with 40~nm height, 46~nm radius and refractive index from Johnson and Christy\cite{Johnson1972} was placed on a substrate with constant refractive index of 1.52. A layer with the templated biaxial HB238 complex refractive index\cite{Schaefer2025} with 30~nm thickness was placed around the particle. A homogeneous refractive index environment was achieved by implementing another layer on top of HB238 with a refractive index of 1.47  corresponding to polyvinyl alcohol. The lattice was simulated by setting periodic boundary conditions (Floquet periodicity) along the substrate surface.  Lattices with periods of 415-535~nm were simulated. 

\section*{Funding sources}

This project is funded with support from the RTG-2591 `TIDE - Template-designed Organic Electronics' (Deutsche Forschungsgemeinschaft). K.L. and R.S. acknowledge funding from the DFG project 426882575. Instrument funding by the Deutsche Forschungsgemeinschaft in cooperation with the Ministerium für Kunst und Wissenschaft of North Rhine-Westphalia (projects 448775637 and INST 216/1121-1 FUGG) is acknowledged.
	  
\section*{Acknowledgements}

Research was supported by the University of Cologne through the Institutional Strategy of the University of Cologne within the German Excellence Initiative (QM$^2$). We thank Tommi Hakala (University of Eastern Finland, Joensuu, Finland) for useful discussions. The help of Philipp Weitkamp (University of Cologne, Cologne, Germany) in developing the molecular templating is acknowledged.\\

\noindent The authors declare no competing financial interest.

\section*{Author contributions}

R.S. fabricated the plasmonic lattices and the samples with a spin-cast HB238 layer, measured and simulated the transmission spectra, and analyzed the data. M.N. synthesized and transferred the graphene nanoribbons, deposited HB238 on the ribbons, and measured and analyzed the atomic force micrographs. K.M. supervised the research carried out by M.N. K.L. planned and supervised the project. R.S. wrote the manuscript with input from all authors.

\providecommand{\latin}[1]{#1}
\makeatletter
\providecommand{\doi}
  {\begingroup\let\do\@makeother\dospecials
  \catcode`\{=1 \catcode`\}=2 \doi@aux}
\providecommand{\doi@aux}[1]{\endgroup\texttt{#1}}
\makeatother
\providecommand*\mcitethebibliography{\thebibliography}
\csname @ifundefined\endcsname{endmcitethebibliography}
  {\let\endmcitethebibliography\endthebibliography}{}

\end{document}